\documentclass[aps,pra,superscriptaddress,twocolumn,showpacs,floatfix]{revtex4-1}

\usepackage{amsmath}
\usepackage{amssymb}
\usepackage{bm}
\usepackage{graphicx}

\begin{document}


\title[]{Photon recoil momentum in a Bose-Einstein condensate of a dilute gas}

\author{Yu. A. Avetisyan}
\affiliation{Precision Mechanics and Control Institute, Russian Academy of Sciences, Saratov 410028, Russia}
\affiliation{Saratov National Research State University, Saratov 410012, Russia}

\author{V. A. Malyshev}
\affiliation{Zernike Institute for Advanced Materials, University of Groningen, Nijnborgh 4, 9747 AG Groningen, The Netherlands}

\author{E. D. Trifonov}
\affiliation{Department of Theoretical Physics, Herzen State University of Russia, St. Petersburg 191186, Russia}


\date{\today}

\begin{abstract}
We develop a "minimal" microscopic model to describe a two-pulse-Ramsay-interferometer-based scheme of measurement of the photon recoil momentum in a Bose-Einstein condensate of a dilute gas [Campbell {\it et al.}, Phys. Rev. Lett. \textbf{94}, 170403 (2005)]. We exploit the truncated coupled Maxwell-Schr\"odinger equations to elaborate the problem. Our approach provides a theoretical tool to reproduce essential features of the experimental results. Additionally, we enable to calculate the quantum-mechanical mean value of the recoil momentum and its statistical distribution that provides a detailed information about the recoil event.
\end{abstract}
%
%
\maketitle

\section{Introduction}
Measurements of the photon momentum in a dispersive medium is of conceptual and practical importance.
Such kind of studies are currently used in quantum metrology, in particular, to determine the ratio $h/m$~\cite{Weiss1993,Battesti2004,Cog2005}, where $h$ and $m$ are the Plank's constant and the atomic mass, respectively, as well as the fine-structure constant $\alpha$~\cite{Wicht2002,Gupta2002}.
In a medium, however, the photon momentum experiences a re-normalization due to the index of refraction, so that the photon momentum $\hbar k_0$ [$\hbar = h/(2\pi)$ and $k_0$ is the vacuum photon wave vector] should be replaced by $n\hbar k_0$, where $n$ is the index of refraction~\cite{Minkovski1908,Haugan1982,Lowden2004,Bradshaw2010,Barnett2010}.

Experimentally, this problem has been tackled by Campbell {\it et al.}~\cite{bib:Campbell_Leanhardt_etal_PheRevLett_94:170403_2005} with a two-pulse light grating (Ramsay) interferometer, using near-resonant laser light.
The scheme of measurements in~\cite{bib:Campbell_Leanhardt_etal_PheRevLett_94:170403_2005} was as follows. An elongated BEC of rubidium atoms $^{87}\rm{Rb}$ in $|5^2S_{1/2} F = 1; m_F = -1\rangle$ state, confined in a magnetic trap, was illuminated in the perpendicular direction with an optical standing wave produced by two identical counter-propagating laser beams of a duration $\delta t$ and of a carrier frequency $\omega_{0}$. As a result, two coherent atomic clouds, moving in the opposite directions, were created. The polarization of the excitation pulses was optimised to suppress the super-radiant Rayleigh scattering in the direction of BEC's elongation. As a result of the Bragg scattering on the optical grating, an atom in its ground state acquires a mean recoil momentum $p$ approximately twice the laser photon momentum $\hbar k_{0} = \hbar\omega_{0}/c$ ($c$ is the speed of light in free space), or recombines to the static cloud. For a given refractive index $n$ of the medium, $p = 2n\hbar k_{0}$. The kinetic energy gained by an atom is equal to $p^{2}/2m$. Therefore, its de Broglie wave frequency is determined by $\omega_{B} = p^{2}/(2m\hbar)$. After some time delay $\tau$, a second identical pulse was applied and the second pair of moving atomic clouds was created. The speed of clouds appears to be low, so that they are not shifted appreciably with respect to each other within the time delay: it leads to their interference and, accordingly, to the density oscillations of clouds as a function of the delay time $\tau$. The latter, in turn, affects the density of the condensate itself, since the total number of atoms is approximately conserved. Measuring the density of the static cloud as a function of the delay time $\tau$ allows one to determine the phase shift $\omega_{B}\tau$ and thus the effective atom recoil momentum. This is, although not a direct, but a highly sophisticated method of measuring the atomic recoil momentum via the influence of the interference of the moving coherent clouds on the condensate itself.

We present a simplified microscopic model of the experiment~\cite{bib:Campbell_Leanhardt_etal_PheRevLett_94:170403_2005} on measuring the photon recoil momentum in a Bose-Einstein condensate of a dilute gas, using a semiclassical theory of the superradiant light scattering (SLS) on a BEC. Within the framework of our approach, we enable, first, to reproduce the essential features of the experiment~\cite{bib:Campbell_Leanhardt_etal_PheRevLett_94:170403_2005} and, additionally, to calculate the quantum-mechanical mean of the recoil momentum in the moving atomic clouds and its statistical distribution.
We demonstrate that the value of the recoil momentum extracted from the interference data for the static cloud~\cite{bib:Campbell_Leanhardt_etal_PheRevLett_94:170403_2005} with a good accuracy coincides with the quantum-mechanical mean. This point on that in the experiment just the latter is measured.

The SLS from a BEC has been observed for the first time in~\cite{bib:Inouye_Science_285:571_1999,bib:Schneble_Science_300:475_2003} and since then an intensive buildup of the theory of the effect has followed~\cite{bib:Moor_Meystre_PhysRevLett_83:5202_1999, bib:Mustecaplioglu_You_PhysRevA_62:063615_2000, bib:Piovella_Gatelli_Bonifacio_OptCommun_194:167_2001, bib:Trifonov_JETP_93:969_2001,  bib:Zhang_Meystre_PhysRevLett_91:150407_2003, bib:Benedek_Benedict_JOptB_6:3_2004, bib:Avetisyan_Trifonov_LaserPhysLett_124_2004_2005_2007,
bib:Robb_Piovella_Bonifacio_JOptB_7:93_2005,
bib:Zobay_Nikolopoulos_PhysRevA_72:041604_2005, bib:Bar-Gill_Rowen_Davidson_PhysRevA_76:043603_2007, bib:Deng_Payne_Hagley_PhysRevLett_104_050402_2010, bib:Avetisyan_Trifonov_PhysRevA_88:025601_2013} that provided substantial insight and understanding of the light-BEC interaction process. More specifically, in Refs.~\cite{bib:Moor_Meystre_PhysRevLett_83:5202_1999,bib:Mustecaplioglu_You_PhysRevA_62:063615_2000,bib:Zhang_Meystre_PhysRevLett_91:150407_2003,
bib:Robb_Piovella_Bonifacio_JOptB_7:93_2005}, the quantum-electrodynamic approach in the mean-field approximation has been used to describe the SLS from a BEC of a dilute cold gas. In papers~\cite{bib:Piovella_Gatelli_Bonifacio_OptCommun_194:167_2001, bib:Trifonov_JETP_93:969_2001,bib:Benedek_Benedict_JOptB_6:3_2004,bib:Zobay_Nikolopoulos_PhysRevA_72:041604_2005, bib:Bar-Gill_Rowen_Davidson_PhysRevA_76:043603_2007,bib:Avetisyan_Trifonov_PhysRevA_88:025601_2013}, the semiclassical approach of the light-matter interaction, naturally incorporating the propagation and nonlinear effects, has been applied to explain essential details of SLS, such as the spatial asymmetry between forward- and backward-moving atomic side modes observed in the strong-pulse regime of SLS~\cite{bib:Zobay_Nikolopoulos_PhysRevA_72:041604_2005}, ultraslow group velocity of the backward-propagating superradiant field~\cite{bib:Deng_Payne_Hagley_PhysRevLett_104_050402_2010}, a crucial role of the multiple recoil processes for SLS on a BEC resulting in that the SLS dominates over the usual Rayleigh scattering~\cite{bib:Avetisyan_Trifonov_PhysRevA_88:025601_2013} and many others.

The paper is organised as follows. In the next section, we present the formalism based on the semiclassical theory of the Ramsay interference in a BEC, involving the coupled system of Maxwell-Schr\"odinger equations within the framework of the slowly-varying amplitude approximation. In Sec.~\ref{Simulations}, the results of simulations of the condensate density oscillations are presented. In Sec.~\ref{Recoil}, we calculate the quantum-mechanical mean value of the recoil momentum and energy of an atom in moving clouds and compare these data with those obtained
in the previous section. Section~\ref{Conclusion} concludes the paper.

\section{Formalism}
\label{Formalism}
In line with the geometry of the experiment~\cite{bib:Campbell_Leanhardt_etal_PheRevLett_94:170403_2005}, we shall use a simplified one dimensional model of the light scattering on a condensate subjected to illumination by two pulses, as described above, separated by a delay time $\tau$. This model underlines the essential features of the problem. An atom will be considered as a two-level Bose-particle with the wave functions $\varphi_a$ and $\varphi_a$ and corresponding eigenenergies $E_{a}$ and $E_{b}$ for the ground and excited states, respectively. We also take into account the atomic translational motion and then seek the atom's wave function in the form
\begin{widetext}
\begin{equation}
\label{eq:psi_xt}
\Psi(x,t) = \\
\sum_{j = 0,\pm 2, \ldots} \left[ \phi_{a,j} a_{j}(x,t)+\exp(-i\omega_{0}t) \phi_{b,j+1} b_{j+1}(x,t)\right]\, ,
\end{equation}
where
\begin{equation}
\label{eq:psi_ab}
\phi_{a,j} = \frac{1}{\sqrt{L}}\exp\left[ik_{0}jx\right]\varphi_{a}\, \quad \mathrm{and} \quad
\phi_{b,j+1} = \frac{1}{\sqrt{L}}\exp\left[ik_{0}(j+1)x\right]\varphi_{b}
\end{equation}
\end{widetext}
are the wave functions of an atom in $|j\rangle$-th and $|j+1\rangle$-th discrete momentum states, respectively, $L$ is the transversal size of the condensate.

To approach the problem, we use the coupled system of the Maxwell-Schr\"{o}dinger (MS) equations and apply the slowly-varying amplitude approximation in time and space. The system of MS equations for amplitudes (in dimensionless units, see below) for our model of the light-condensate interaction reads
\begin{widetext}
\begin{subequations}
\begin{equation}
\label{eq:ms_aj}
\frac{\partial a_{j}(x,t)}{\partial t} + v_{j}\frac{\partial a_{j}(x,t)}{\partial x} = -i\omega_{j}a_{j}(x,t) + \bar{E}^{+}b_{j+1}(x,t) + \bar{E}^{-}b_{j-1}(x,t)\, ,
\end{equation}
\begin{eqnarray}
\label{eq:ms_bj}
\frac{\partial b_{j+1}(x,t)}{\partial t} + v_{j+1}\frac{\partial b_{j+1}(x,t)}{\partial x} \nonumber \\
= i\left(\Delta -\omega_{j+1} + i \gamma/2\right)b_{j+1}(x,t) - E^{+}a_{j}(x,t) - E^{-}a_{j+2}(x,t)\, ,
\end{eqnarray}
\begin{equation}
\label{eq:E_plus}
E^{+}(x,t) = E_{0}(t) + 2\int^{x}_{0}dx'\sum_{j = 0,\pm 2,\ldots}b_{j+1}(x',t)\bar{a}_{j}(x',t)\, ,
\end{equation}
\begin{equation}
\label{eq:E_minus}
E^{-}(x,t) = E_{0}(t) + 2\int^{1}_{x}dx'\sum_{j = 0,\pm 2,\ldots}b_{j-1}(x',t)\bar{a}_{j}(x',t)\, ,
\end{equation}
\end{subequations}
\end{widetext}
where $j = 0, \pm 2, \pm 4, \ldots$  We adapted in Eqs.~\eqref{eq:ms_aj} - \eqref{eq:E_minus} as units of length and time, respectively, the condensate transversal size $L$ and the superradiant time constant $\tau_{R} = \hbar/(\pi d^2 k_{0}N_{0}L)$~\cite{bib:Benedict_Super-radiance_1996}, where $d$ is the atom transition dipole moment and $N_{0}$ is the atom number density. The slowly varying field amplitudes of the forward (backward) $E^{+}$ ($E^{-}$) and incident $E_{0}$ fields are scaled by $i\hbar/(d\tau_{R})$ (overbars denote the complex conjugation). The quantities $\omega_{j} = \hbar j^{2}k^{2}_{0}\tau_{R}/(2m)$ and $v_{j} = j\hbar k_{0}\tau_{R}/mL$ are the dimensionless atom recoil frequency and velocity, respectively, where the index $j$  runs over $0,\pm 2, \pm 4, \ldots$ for the ground state, while over $\pm 3, \pm 5, \ldots$ for the excited state. Furtheremore, $\Delta = (\omega_{0}-\omega_{ba})\tau_{R}$ is the dimensionless detuning of the incident field frequency $\omega_{0}$ away from the atomic resonance at $\omega_{ba}$, $\gamma = \Gamma\tau_{R}$, where $\Gamma$ is the spontaneous emission rate of the excited atomic state. The retardation is neglected in Eqs.~(\ref{eq:E_plus}) and~(\ref{eq:E_minus}) as the flight time of light through the system $L/c$ is much shorter than all other times in the problem. The only nonzero initial condition to the system of equations~\eqref{eq:ms_aj} - \eqref{eq:E_minus} is $a_{0}(x,t=0) = 1$. All other variables equal to zero before the first excitation pulse arrives.

The similar system of equations~(\ref{eq:ms_aj}) - (\ref{eq:E_minus}) has been previously used for the description of the super-radiant scattering on BECs of a dilute gas~\cite{bib:Avetisyan_Trifonov_PhysRevA_88:025601_2013}, only without spatial derivatives of the wave function amplitudes $a_j(x,t)$ and $b_j(x,t)$. While those terms has no effect on the final results in the underlined studies, in our case, they are of crucial importance to catch out some fine features of the two-pulse Ramsay interference (see for details the next section).
Additionally, we do not use the approximation of adiabatic elimination of the excited atomic state, usually assumed when considering the light-condensate interaction. This allows us to consider an arbitrary value of the detuning $\Delta$, i.e. to scan the exact resonance, $\Delta = 0$, that is important for our study.

\section{Condensate density oscillations}
\label{Simulations}
In our modeling of the Ramsay interference, we used the set of the system's parameters, approximated to those in the experiment~\cite{bib:Campbell_Leanhardt_etal_PheRevLett_94:170403_2005}:
the BEC transversal size $L = 16\,\mu$m, the atom number density $N_{0} = 4.15\times 10^{13}$cm$^{-3}$, the radiation constant of the $|5^2P_{3/2} F = 1\rangle \rightarrow |5^2S_{1/2} F = 1\rangle$ transition (the wavelength $\lambda = 780$~nm) $\Gamma = 0.37 \cdot 10^8$ s$^{-1}$, the corresponding transition dipole moment $d = 2.07\cdot 10^{-29}$~C\,m. For these parameters, the super-radiant constant is estimated to be $\tau_{R} \approx 1.75\times 10^{-9}$~s. Then, for the dimensionless quantities $v_j$, $\omega_j$ and $\gamma$ one gets: $v_{j} = 7.8\times 10^{-7}\,j$, $\omega_{j} = 5\times 10^{-5}\,j^{2}$, $\gamma = 5\times 10^{-2}$. The detuning $\Delta$ was varied within a range of $[-12,12]$.

To excite the condensate, we used rectangular pulses of duration $\delta t = 5\mu$s ($\delta t/\tau_R = 3\times 10^{3}$). The delay time $\tau$ between the pulses was varied within a range of [5,150]~$\mu$s (in dimensionless units $\tau/\tau_R$, within $[3,90] \times 10^{3}$). The dimensionless amplitude of the incident pulse, $E_{0} = 6\times 10^{-3}$ was chosen so that after a $\delta t$-long excitation, the population of the static cloud decreased approximately to a level of 0.9. When solving the MS equations~\eqref{eq:ms_aj} - \eqref{eq:E_minus}, we took into account the generation of atomic clouds up to the 10-th order.

\begin{figure}[ht!]
\begin{center}
\includegraphics[width=0.476\columnwidth]{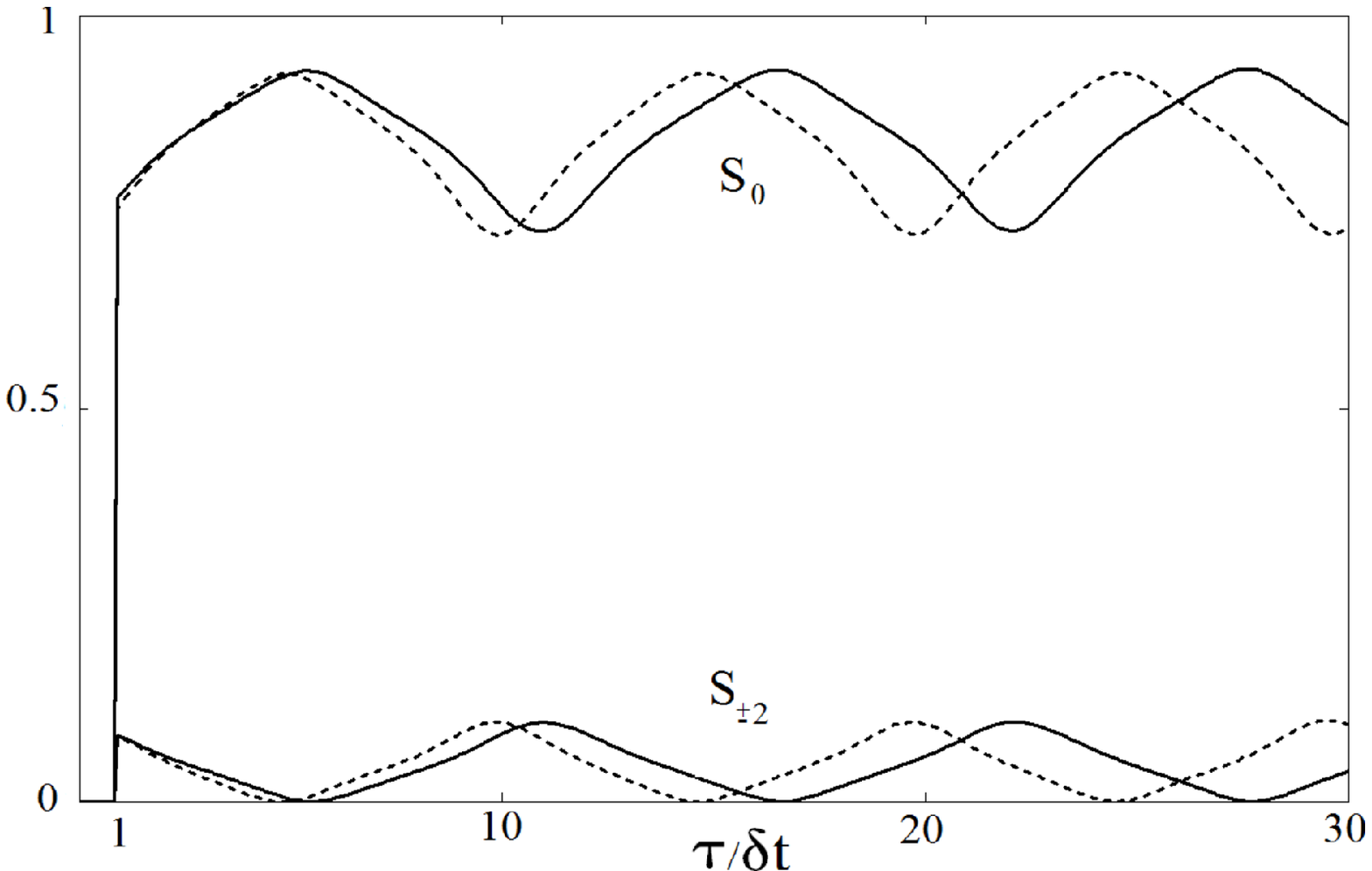}
\includegraphics[width=0.465\columnwidth]{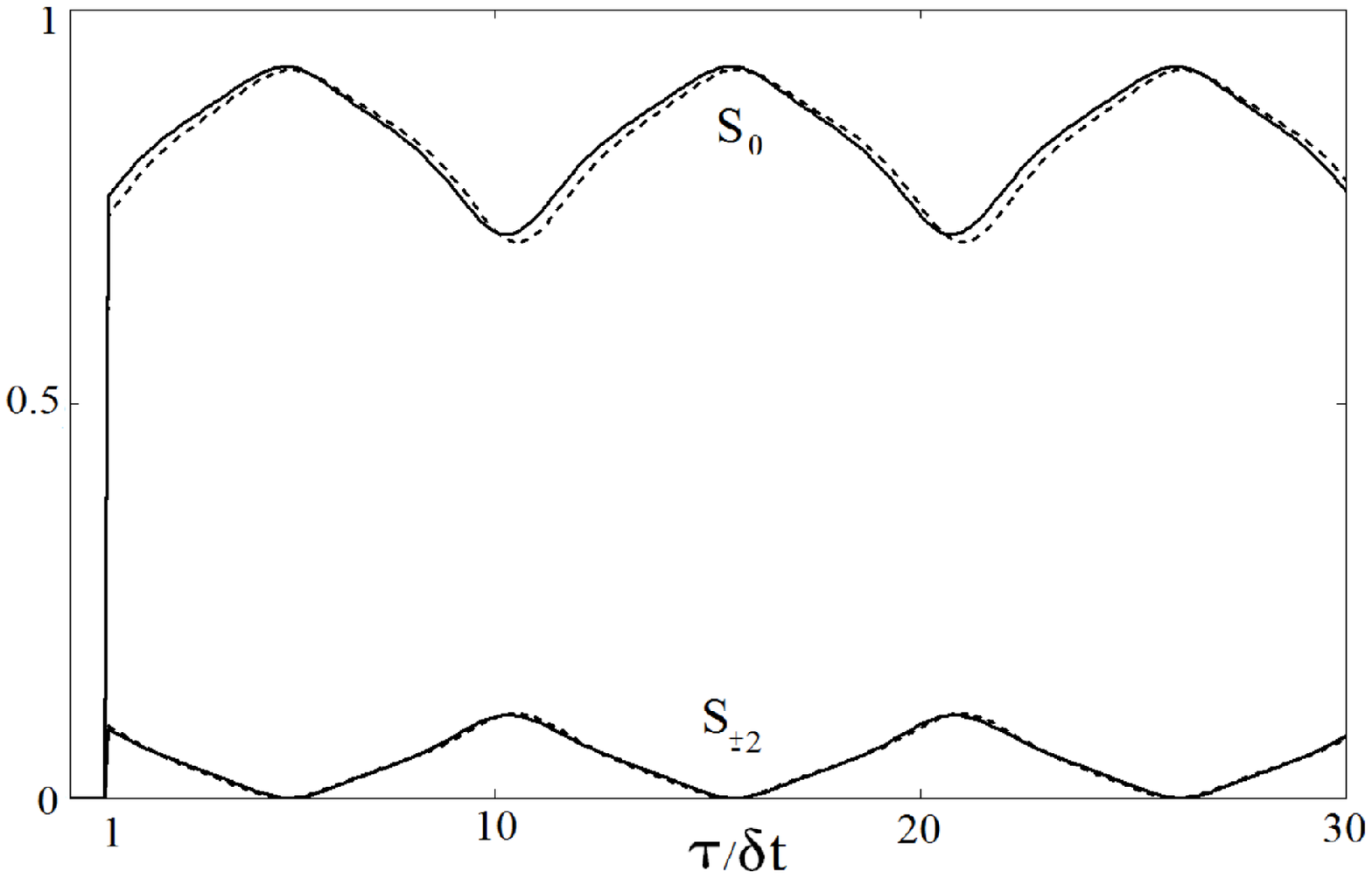}
\end{center}
\caption{\label{fig:IntAtClouds} Lefft panel - Interference fringes of $S_{0}(\tau + \delta t)$ and $S_{\pm 2}(\tau + \delta t)$ for $|0\rangle$ - and $|\pm 2\rangle$ - clouds, respectively. The solid (dashed) curve was calculated for $\Delta = 0.5$ ($\Delta = -0.5$).
Right panel - same, only neglecting the first spatial derivatives of the wave function amplitudes $a_j(t)$ and $b_j(t)$ in Eqs.~(\ref{eq:ms_aj}) and~(\ref{eq:ms_bj}).}
\end{figure}

First of all, we are interested in the fraction of atoms in the static BEC cloud at a time point $t = \tau + \delta t$, when the second excitation pulse has gone. It is defined as
\begin{equation}
\begin{split}
\label{eq:s0}
S_{0}(t) = \int^{1}_{0}dx \left|a_{0}(x,t)\right|^{2} \, .
\end{split}
\end{equation}
Also, it is of our interest the fraction of atoms in the moving atomic clouds in the ground state, $|\pm 2j\rangle$, $j = 1,2,3,\ldots$:
\begin{equation}
\begin{split}
\label{eq:spm2j}
S_{\pm 2j}(t) = \int^{1}_{0} dx \left|a_{\pm 2j}(x,t)\right|^{2} \, .
\end{split}
\end{equation}
The results of simulations for the static $|0\rangle$ and moving $|\pm 2\rangle$ clouds of BEC are shown in Fig.~\ref{fig:IntAtClouds}. As is seen from the figure, the fraction of atoms in all clouds, $S_0(t)$ and $S_{\pm 2}$, as a function of the delay time $\tau$,  reveals oscillations as it has been observed in the experiment~\cite{bib:Campbell_Leanhardt_etal_PheRevLett_94:170403_2005}.
The authors of~\cite{bib:Campbell_Leanhardt_etal_PheRevLett_94:170403_2005} associated the density oscillations of the static condensate cloud mainly with those of the $|\pm 2\rangle$ clouds. This finds its confirmation in our calculations.
Indeed, one observes strong correlations in frequencies, phases and amplitudes of the $S_{\pm 2}$ and $S_{0}$ interference fringes for the same $\Delta$, that reflects the conservation of the total number of atoms in these states. We emphasize that interference fringes are shifted with respect to each other for different signs of $\Delta$. The most important, however, is that their frequencies deviate from each other for altered signs of the detuning $\Delta$ that agrees with the experiment~\cite{bib:Campbell_Leanhardt_etal_PheRevLett_94:170403_2005}. The underlined shift disappears, if one neglects in Eqs.~\eqref{eq:ms_aj} and~\eqref{eq:ms_bj} the terms proportional to the spatial derivative of the wave function amplitudes $a_j(t)$ and $b_j(t)$ (see Fig.~\ref{fig:IntAtClouds}, right panel). This points on the relevance of those terms for the correct description of the Ramsay interference.

As is seen from Fig.~\ref{fig:IntAtClouds}, the density oscillations do not show any decay which has been found in Ref.~\cite{bib:Campbell_Leanhardt_etal_PheRevLett_94:170403_2005} and has been explained there by decreasing the overlap between the recoiling atoms and those at rest due to the motion away after the shutoff of the magnetic trap. In our case, the system is confined in a finite interval $[0, L]$, i.e, there is no expansion of clouds and, consequently, any decay of the interference signal.

From the density oscillations obtained, one can extract the recoil frequency $\omega_{rec}$ for different values of the detuning $\Delta$.
Analyzing the numerical data presented in the left plot of Fig.~\ref{fig:IntAtClouds}, we found that $\omega_{rec} \approx 1.06\, \omega_{2}$ for $\Delta = -0.5$, while $\omega _{rec} \approx 0.94 \, \omega_{2}$ for $\Delta = 0.5$, where $\omega_2 = 4\hbar k^{2}_{0}\tau_{R}/(2m)$ is the bar frequency of atoms in the $|\pm 2\rangle$ coherent clouds. We point out again that the recoil frequency $\omega_{rec}$ differs for altered signs of the detuning $\Delta$. Oppositely, the similar analysis performed for the right plot of Fig.~\ref{fig:IntAtClouds} yields $\omega_{rec} = \omega_2$ independently of the sign of $\Delta$.

\section{Momentum and frequency recoil}
\label{Recoil}
The density oscillations of the static cloud of the condensate provides a tool to determine the actual value of the atom recoil momentum/frequency in moving clouds, as has been implemented in~\cite{bib:Campbell_Leanhardt_etal_PheRevLett_94:170403_2005} within the framework of the phenomenological picture. Our microscopic approach allows one to get a more detailed information about the recoil momentum/frequency for different atomic clouds: not only its mean value, but also the distribution function. The latter for the $j$-th atomic cloud is defined as
\begin{equation}
\begin{split}
\label{eq:wj}
w_{j}(k,t) = \frac{\left|f_{j}(k,t)\right|^{2}}{\int^{+\infty}_{-\infty} dk^{\prime}\left|f_{j}(k',t)\right|^{2}}\ ,
\end{split}
\end{equation}
where $f_{j}(k,t) = \int^{1}_{0}dx\exp\left(-ikx\right)a_{j}(x,t)$ is the Fourier transform of the amplitude $a_{j}(x,t)$.
\begin{figure}[ht!]
\begin{center}
\includegraphics[width=0.47\columnwidth]{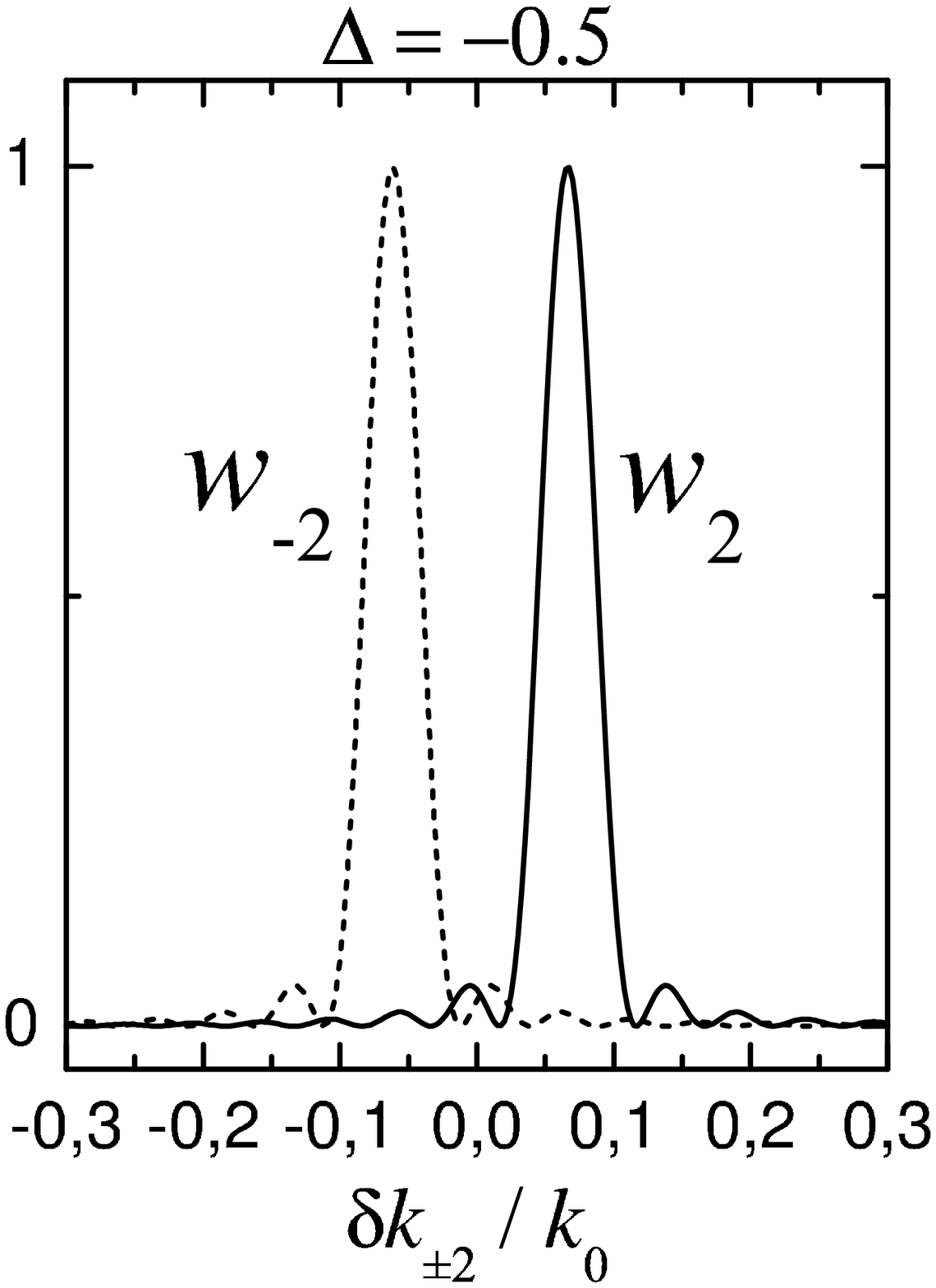}
\includegraphics[width=0.47\columnwidth]{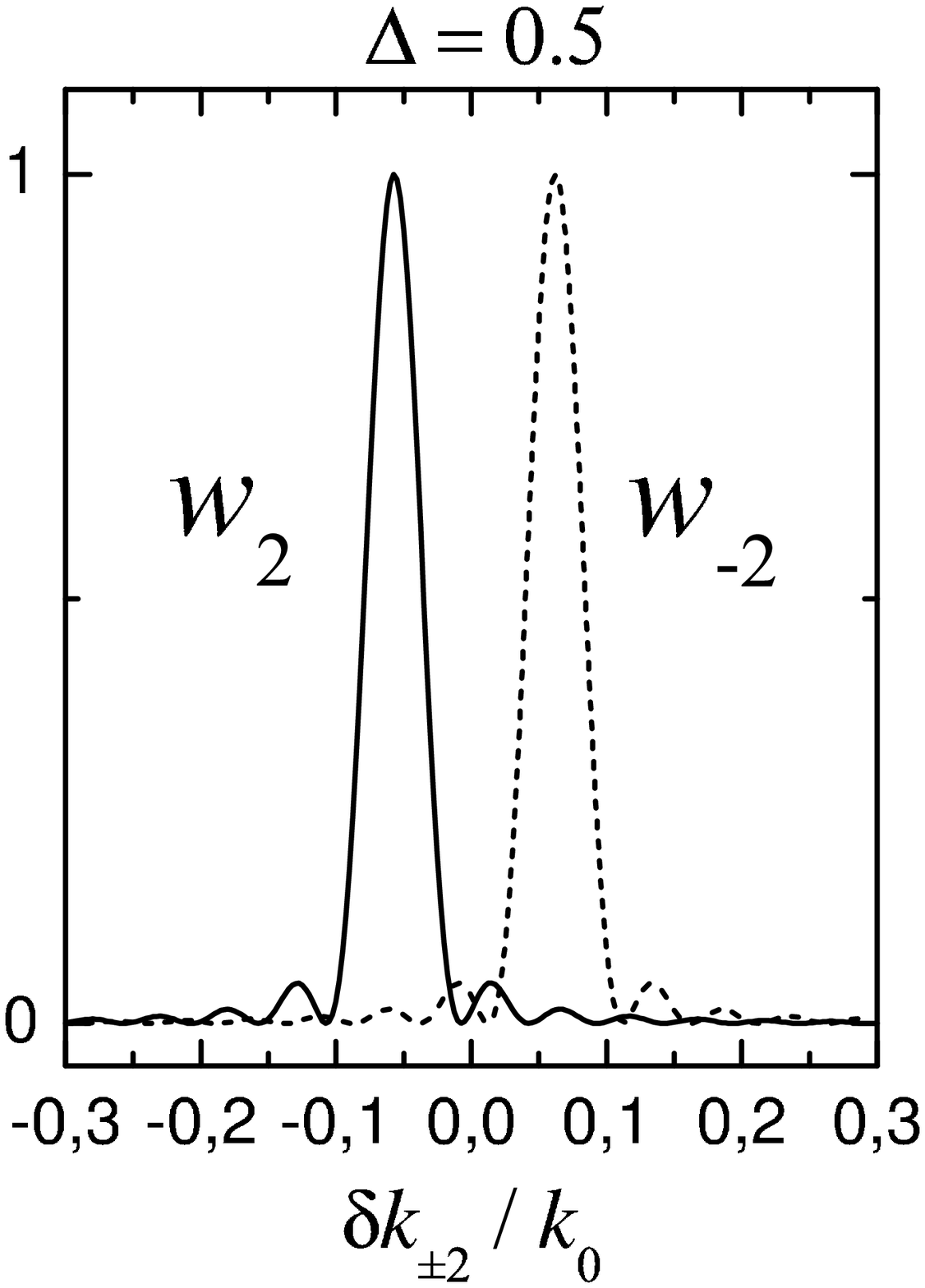}
\end{center}
\caption{\label{fig:DistribFuncRecMoment}
Distribution functions of the recoil momentum shift $\delta k_{\pm 2} = k_{\pm 2} - 2k_0$ for atoms in the $|-2\rangle$ and $|2\rangle$ clouds (dashed and solid curves, respectively) immediately after the first pulse excitation of $\delta t = 5\mu$s - duration for two values of the detuning: $\Delta = -0.5$ (left panel) and $\Delta = 0.5$ (right panel).}
\end{figure}
Then for the mean recoil momentum of an atom in the $|j\rangle$-th cloud one gets
\begin{equation}
\begin{split}
\label{eq:deltak}
k_{j} = \int^{+\infty}_{-\infty} dk \,k w_{j}(k,t)\, ,
\end{split}
\end{equation}
and for its variance:
\begin{equation}
\begin{split}
\label{eq:Dj}
D_{j} = \int^{+\infty}_{-\infty} dk\, w_{j}(k,t)(k - k_{j})^{2} \, .
\end{split}
\end{equation}

Examples of the distribution functions for the recoil momentum shift $\delta k_{\pm 2} = k_{\pm 2} - 2k_0$ of atoms in the $|\pm 2\rangle$ clouds, obtained immediately after the action of the first pulse, are shown in Fig.~\ref{fig:DistribFuncRecMoment}.
As is seen, the distribution for $|2\rangle$ ($|-2\rangle$) clouds is sign-dependent (symmetric with respect to $\Delta = 0$), which is coherent with the data of the interference fringes. Thus, the Fourier transform of the signal after the action of the fist pulse already contains the information deduced from the interference fringes, i.e. after the action of the second pulse.

\begin{figure}[ht!]
\begin{center}
\includegraphics[width=0.9\columnwidth]{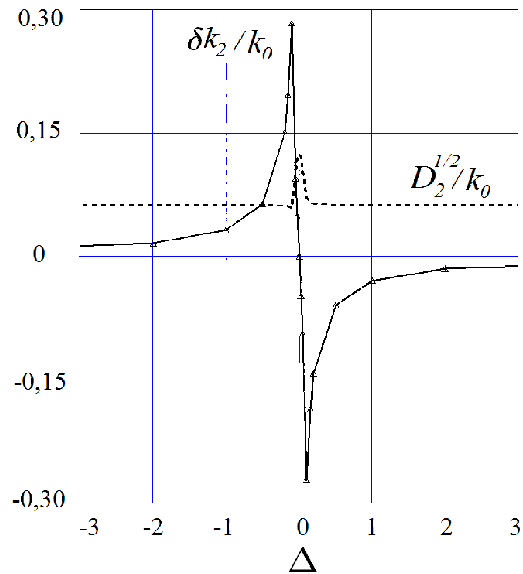}
\end{center}
\caption{\label{fig:Momentum_Shift}
The mean recoil momentum shift $\delta k_{2} = k_{2} - 2k_0$ (solid curve) and its standard deviation $D_2^{1/2}$ (dashed curve), both in units of $k_{0}$, for an atom in the $|2\rangle$ cloud versus the detuning $\Delta$.}
\end{figure}

The results for the mean recoil momentum shift $\delta k_2 = k_2 - 2k_0$ and its standard deviation $D_2^{1/2}$ as a function of the detuning $\Delta$ for an atom in the $|2\rangle$ cloud are depicted in Fig.~\ref{fig:Momentum_Shift}.
Note that $\delta k_2 = k_2 - 2k_0 = 2k_0(n - 1)$, and thus Fig.~\ref{fig:Momentum_Shift} represents in fact the detuning dependence of the refraction index $n$.
The $\Delta$ - dependence of $\delta k_{-2} = k_{-2} - 2k_0$ for the $|-2\rangle$ cloud is mirror-symmetric with respect to that of $\delta k_2$.

From Fig.~\ref{fig:Momentum_Shift}, one can see that changing the sign of the detuning $\Delta$ alters the sign of the mean recoil momentum shift $\delta k_{2}$. Because of that, the $\Delta$ - dependence of $\delta k_{2}$ has a dispersive shape. We point out on a relatively large standard deviation $D_2^{1/2}$ of $\delta k_{2}$. This is a result of the finite BEC's size $L$, as well as the spatial inhomogeneity of the atomic state amplitude, representing the main source of uncertainty of the recoil momentum.

\begin{figure}[ht!]
\begin{center}
\includegraphics[width=0.5\columnwidth]{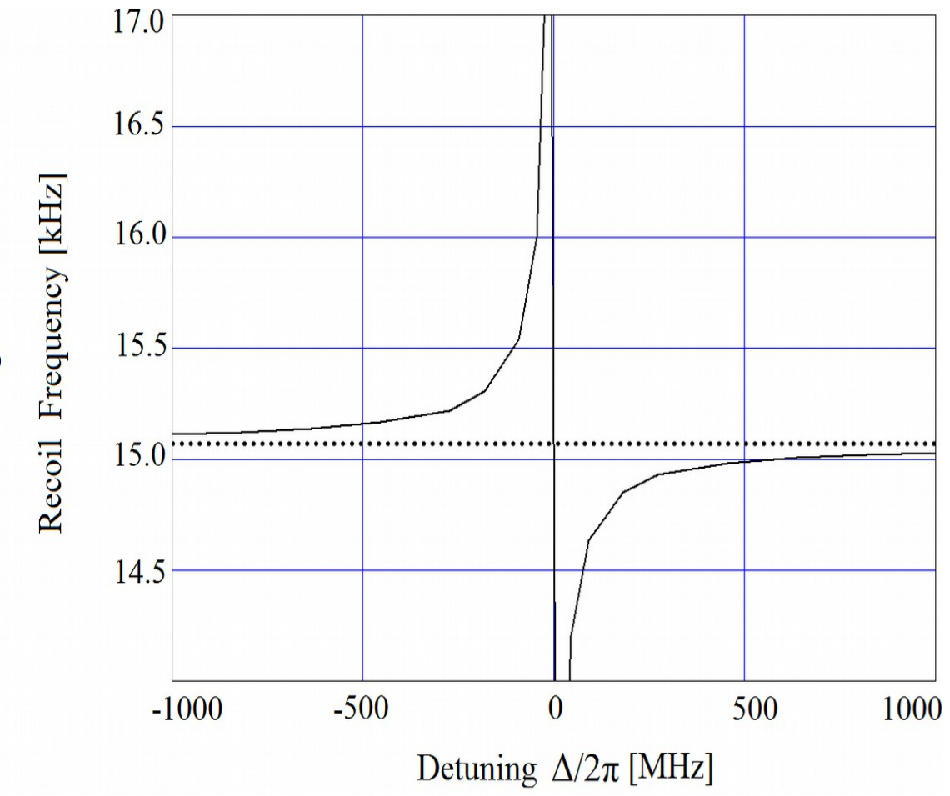}
\includegraphics[width=0.485\columnwidth]{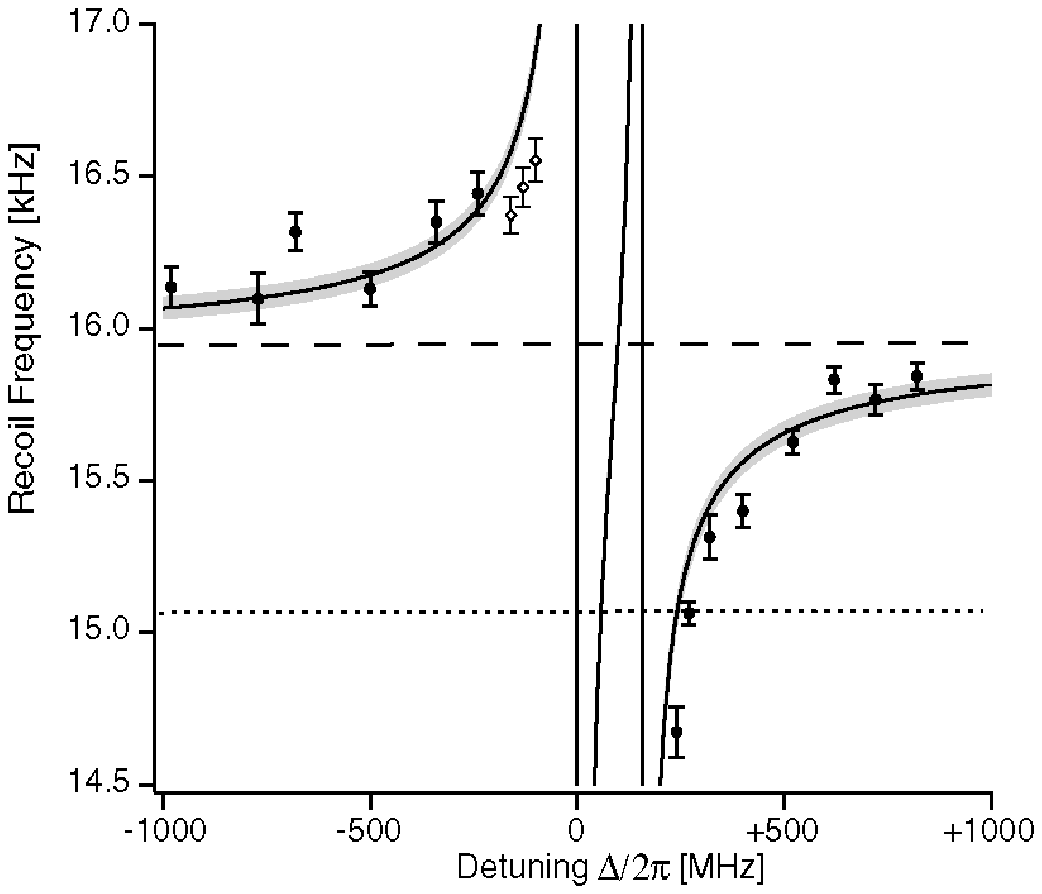}
\end{center}
\caption{\label{fig:Recoil frequency}
The calculated detuning dependence of the recoil frequency $\omega_{rec}$ (left plot) and the measured one (right plot), taken from~\cite{bib:Campbell_Leanhardt_etal_PheRevLett_94:170403_2005}, Fig. 3. The dotted lines show the two-photon recoil frequency $4\omega_{rec} = 15 068$ Hz.
For further explanation, see text and Ref.~\cite{bib:Campbell_Leanhardt_etal_PheRevLett_94:170403_2005}, Fig. 3.
}
\end{figure}

We recalculated the data presented in Fig.~\ref{fig:Momentum_Shift} into the recoil frequency $\omega_{rec}$, Fig.~\ref{fig:Recoil frequency} (left plot). For comparison, in Fig.~\ref{fig:Recoil frequency} (right plot) the experimental results of~\cite{bib:Campbell_Leanhardt_etal_PheRevLett_94:170403_2005} for $\omega_{rec}$ are shown. Contrasting these two plots, we see that the theoretical and experimental curves (thick dots with error bars) have in common the dispersive shape of the $\Delta$-dependence of the recoil frequency. However, the experimental curve has two features that distinguish it from the theoretical one. First, it is shifted up by approximately 900 KHz (dashed line) that is due to the so-called mean-field shift~\cite{bib:Campbell_Leanhardt_etal_PheRevLett_94:170403_2005}), and second, the lower part of it is displaced to the right by 157 MHz, because of the presence of the other allowed transition $|5^2S_{1/2} F = 1\rangle \rightarrow |5^2P_{3/2} F = 2\rangle $, contributing to the optical response~\cite{bib:Campbell_Leanhardt_etal_PheRevLett_94:170403_2005}. These two effects are not taken into account in our simplified theory. The most important fact is that our approach recovers the dispersive shape of the detuning dependence of the recoil frequency $\omega_{rec}$.

Now, let us compare the result for the quantum-mechanical mean recoil frequency shift $\delta\omega_2$ with the one obtained in the simulations of the Ramsay interference, $\delta\tilde{\omega}_2$. We use for that a relation $\delta\omega_{2}/\omega_{2} = 2\delta k_{2}/k_{2}$ between the frequency shift $\delta\omega_{2}$ and the momentum shift $\delta k_{2} = k_2 - 2k_0$. From the calculated for $\delta k_2$ data we obtained $\delta\omega_{2}/\omega_{2}\approx 0.062$ for $\Delta = -0.5$ and $\delta\omega_{2}/\omega_{2}\approx -0.057$ for $\Delta = 0.5$. At the same time, from the interference fringes (Fig.~\ref{fig:IntAtClouds}, left panel), the corresponding values are found to be $\delta\tilde{\omega}_{2}/\omega_{2}\approx 0.059$ and $\delta\tilde{\omega}_{2}/\omega_{2}\approx -0.058$, respectively. A comparison of these data shows that with a good accuracy the value of the recoil frequency shift $\delta\omega_2$, extracted from the interference fringes, coincides with
the quantum-mechanical mean. This point on that in the experiment just the the latter quantity is measured.

\section{Conclusion and outlook}
\label{Conclusion}
We have presented a microscopic theory, reproducing the essential features of the experimental results on measuring the photon recoil momentum in a BEC of a dilute gas by means of the two-pulse Ramsay interference~\cite{bib:Campbell_Leanhardt_etal_PheRevLett_94:170403_2005}.
For this purpose, we have used the coupled Maxwell-Schr\"odinger equations within the framework of the slowly-varying envelope approximation. We have found that for the adequate description of the experiment~\cite{bib:Campbell_Leanhardt_etal_PheRevLett_94:170403_2005}, it is of principal importance to take into account corrections to the bare recoil energy of an atom because of the inhomogeneity of atomic clouds (the spatial derivatives of the atom wave function amplitudes). Neglecting them results (in the theory) in that the photon recoil momentum in the medium coincides with its vacuum value. The microscopic approach has allowed us to directly calculate the quantum-mechanical mean value of the recoil momentum of an atom and its statistical distribution in moving atomic clouds. We have found that the recoil momentum, extracted from the interference fringes of the static BEC cloud, as it has been done in the experiment~\cite{bib:Campbell_Leanhardt_etal_PheRevLett_94:170403_2005}, represents just the quantum-mechanical mean value.

We have considered a Bose-Einstein condensate of an ideal atomic gas. A question that remains to answer is to what extent the interaction between atoms (within the microscopic picture) will affect the Ramsay interference?  Additionally, we have used in our analysis the slowly varying amplitude approximation in space for fields. Keeping the second space derivative in the Maxwell equations will allow one to correctly take into account the reflection of the laser beams from the condensate as well as the fields inside the condensate from the boundaries of the latter. The effects of diffraction of beams on the Ramsay interference is also a question to be answered. These issues are a subject of a forthcoming paper.

\acknowledgments
The authors would like to thank M. G. Benedict, A. K. Belyaev and V. V. Tuchin for discussions and I. V Ryzhov for technical assistance.
E. D. T. acknowledges support from the Russian Foundation for Basic Research (grant~15-02-08369 -A). Yu. A. A. thanks the Russian Scientific Foundation (grant 16-19-10455) for support in a part of developing advanced algorithms for the analysis of the primary photon-single-scatter event.

\end{document}